\documentclass{article}
\usepackage{dcase2018,amsmath,graphicx,url,times,booktabs, tabularx,pdfpages,multirow,caption}
\usepackage{arydshln} 
\usepackage{units}
\usepackage{float}

\title{DCASE 2018 Challenge - Task 5: \\Monitoring of domestic activities based on multi-channel acoustics}

 \name{Gert Dekkers$^{1,2,}\sthanks{Thanks to VLAIO-SBO project Sound INterfacing through the Swarm (SINS) \cite{SINS} for funding (contract 130006).}$,
		Lode Vuegen$^{1,2}$,
		Toon van Waterschoot$^{2}$,
		Bart Vanrumste$^{2}$,
		Peter Karsmakers$^{1}$
       }
 \address{$^1$ KU Leuven, Department of Computer Science, Geel, Belgium.\\
		$^2$ KU Leuven, Department of Electrical Engineering, Leuven, Belgium.\\
  }

\begin{document}

\ninept
\maketitle

\begin{sloppy}

\begin{abstract}
The DCASE 2018 Challenge consists of five tasks related to automatic classification and detection of sound events and scenes. This paper presents the setup of Task 5 which includes the description of the task, dataset and the baseline system. In this task, it is investigated to which extent multi-channel acoustic recordings are beneficial for the purpose of classifying domestic activities. The goal is to exploit spectral and spatial cues independent of sensor location using multi-channel audio. For this purpose we provided a development and evaluation dataset which are derivatives of the SINS database and contain domestic activities recorded by multiple microphone arrays. The baseline system, based on a Neural Network architecture using convolutional and dense layer(s), is intended to lower the hurdle to participate the challenge and to provide a reference performance.
\end{abstract}

\begin{keywords}
Acoustic scene classification, Multi-channel, Activities of the Daily Living 
\end{keywords}

\section{Introduction}
\label{sec:intro}
There is a rising interest in smart environments that enhance the quality of life for humans in terms of e.g. safety, security, comfort, and home care \cite{Erden2016}. 
In order to have smart functionality, situational awareness is required, which might be obtained by interpreting a multitude of sensing modalities including acoustics. Compared to other modalities, microphone sensors contain highly informative data which can be exploited for multiple purposes \cite{Vacher2011A}. However, many challenges remain regarding the automatic recognition of sounds. In order to properly compare different computational methods the community needs common publicly available datasets. Previous editions of the Detection and Classification of Acoustic Scenes and Events (DCASE) challenge offered a competitive platform to compare different computational methods using common datasets for various problems related to automatic classification and detection of sound events and scenes \cite{DCASE2017,DCASE2013,DCASE2016}. This year's challenge, DCASE 2018 \cite{DCASE2018Challenge}, consists of five tasks. This paper describes Task 5 which resides in the context of Ambient Assisted Living (AAL) where persons are monitored, e.g. to support patients with a chronic illness and older persons, by tracking their activities being performed at home \cite{Vacher2011A, VuegenL2013A,VuegenL2015A,Vacher2014}. When considering an acoustic sensing modality, a domestic activity can be seen as an acoustic scene. An acoustic event is defined as a single consecutive event originating from a single sound source, e.g. a hand clap or a door knock. 
The ensemble of multiple events create an acoustic scene describing a certain environment (e.g. a park or a living room) or, relevant to this task, an activity being performed by a person (e.g. cooking or watching TV). The acoustic sensing literature has mainly covered the problems of automatic classification and detection of sound events and scenes by using spectral information \cite{VuegenL2013A,VuegenL2015A,DCASE2017}. Similarly, previous DCASE challenges did not focus on letting participants exploit spectral and spatial information. Task 5 offers a multi-channel dataset to compare computional methods that use both types of information. The goal is to exploit spectral and spatial cues independent of sensor location using multi-channel audio for the purpose of classifying domestic activities. 
\\This paper presents DCASE 2018 Task 5 in detail. A task definition, information about the dataset, the task setup, baseline system, and baseline results on the development dataset are provided.

\section{TASK SETUP}
\label{sec:taskdesc}
\subsection{Description}
\label{subsec:desc}
The goal of this task is to classify multi-channel audio segments (i.e. segmented data is given), acquired by a microphone array, into one of the provided predefined classes as illustrated by Figure \ref{fig:taskdesc}. These classes are daily activities performed in a home environment (e.g. “Cooking”, “Watching TV” and “Working”). 

\begin{figure}[h]
  \centering
  \centerline{\includegraphics[width=0.55\columnwidth]{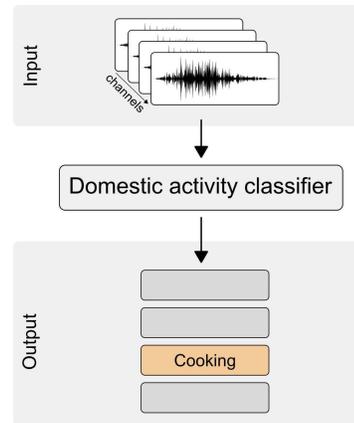}}
  \caption{DCASE 2018 Task 5 description}
  \label{fig:taskdesc}
\end{figure}

As they can be composed out of different sound events, such activities are considered as acoustic scenes. Therefore, this task is quite similar to \textit{DCASE~2018~Task~1:~Acoustic~scene~classification}. The difference lies in the type of scenes and the possibility to use multi-channel audio.
\\For the given problem a single person living at home is considered. This reduces the complexity since the number of overlapping activities is expected to be small. In fact, in the considered data set no overlapping activities are present. These conditions were chosen to be able to focus on the main goal of this task which is to investigate to which extent multi-channel acoustic recordings are beneficial for the purpose of detecting domestic activities. This means that spatial properties can be exploited to serve as input features to the classification problem. However, using estimates of absolute angles or positions of sound sources as input for the detection model is doomed not to generalize well to cases where the position of the microphone array is altered. Therefore, in this task the focus is on systems which can exploit spatial cues independent of sensor position using multi-channel audio.

\subsection{Timeline}
\label{subsec:timeline}

The DCASE 2018 Challenge consists of five tasks related to automatic classification and detection of sound events and scenes \cite{DCASE2018Challenge}. All these tasks follow the same timeline and similar submission guidelines. Table \ref{table:taskplan} introduces the timeline of the Challenge/Task. First, a development dataset was provided along with reference annotations and a baseline system. A month prior to the submission deadline the evaluation dataset was released. Challenge submissions consist of a systems output on the evaluation dataset formatted according to the requirements described in \cite{DCASE2018Task5}. In order to compare and understand all system(s), participants were also required to submit a technical report containing the description of the system(s) in detail. Reference annotations for the evaluation data were only available to us, therefore we were responsible for evaluating the results according to the defined metric \cite{DCASE2018Task5}. These results will be made public after evaluation. Finally, the results will also be presented on the DCASE2018 Workshop. Participants could optionally submit their technical report as a paper to this conference. 

\begin{table}[t]
\centering
\begin{tabular}{c|c} 
Release of development dataset		& 30 March 2018\\ 
Release of baseline system		& 16 April 2018\\ 
Release of evaluation dataset		& 30 June 2018\\ 
Challenge submission			& 31 July 2018\\ 
Publication of results			& 15 Sept 2018\\
DCASE2018 Workshop			& 19-20 Nov 2018\\  
\end{tabular}
  \caption{Challenge/Task timeline}
  \label{table:taskplan}
\end{table}

\begin{figure}[b]
  \centering
  \centerline{\includegraphics[width=0.6\columnwidth]{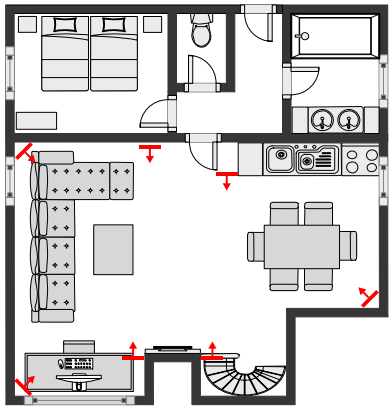}}
  \caption{2D map of the vacation home}
  \label{fig:vachome}
\end{figure}

\subsection{Development and evaluation dataset}
\label{subsec:devdataset}

The datasets used in this task are a derivative of the SINS dataset \cite{Dekkers2017}. The SINS dataset contains a continuous recording of one person living in a vacation home over a period of one week. It was collected using a network of 13 microphone arrays distributed over the entire home. The microphone array consists of four linearly arranged microphones. For this task seven microphone arrays in the combined living room and kitchen area are used. More information about the SINS dataset can be found in \cite{Dekkers2017}. Figure \ref{fig:vachome} shows the floorplan of the recorded environment along with the position of the used microphone array.

\begin{figure}[t]
  \centering
  \centerline{\includegraphics[width=0.55\columnwidth]{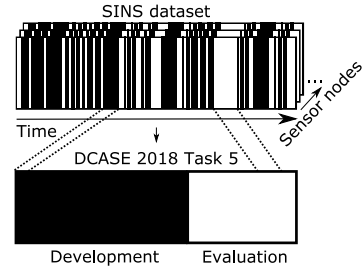}}
  \caption{From SINS to DCASE 2018 Task 5 dataset}
  \label{fig:data}
\end{figure}

\begin{figure*}
  \centering
  \centerline{\includegraphics[width=2\columnwidth]{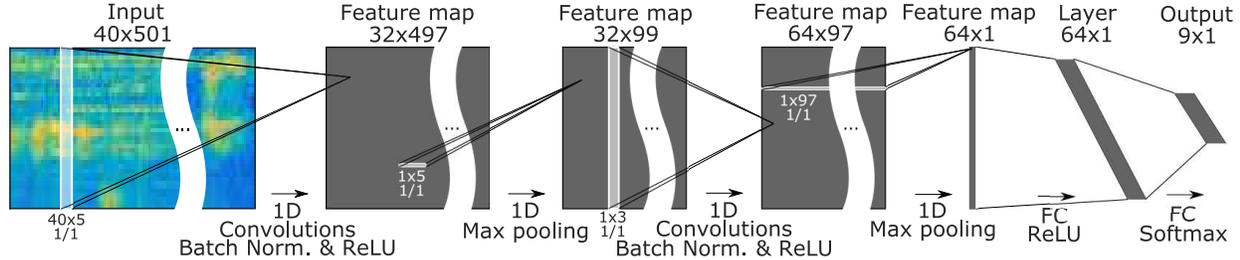}}
  \caption{Neural Network architecture of the baseline system consisting of two 1D convolutional layers and one fully connected layer}
  \label{fig:nn}
\end{figure*}

For both the development and evaluation dataset the continuous recordings were split into audio segments of 10 s. We choose this window size as it is in the same order as the shortest duration of an activity.  Each audio segment contains four channels which represent the four microphone channels from a particular microphone array. Segments containing more than one active class (e.g. a transition of two actitivies) were left out. This means that each segment represents a single activity.  For the development and evaluation set we took a time-wise subset (from the SINS dataset) of \nicefrac{2}{3} and \nicefrac{1}{3} respectively. Full sessions of a particular activity were kept together as shown by Figure \ref{fig:data}. Approximately 50\% of the data was left out by subsampling to make the dataset easier to use for a challenge. 

\begin{table}[b]
\centering
\begin{tabular}{cccc} 
\toprule
 Set & Activity & Segments & Sessions\\
\midrule
\multirow{10}{*}{\rotatebox[origin=c]{90}{Development}} 
 & Absence 				& 18860 			& 42\\ 
 & Cooking				& 5124 			& 13\\  
 & Dishwashing   			& 1424 			& 10\\ 
 & Eating 				& 2308			& 13\\
 & Other 				& 2060 			& 118\\  
 & Social Activity 			& 4944 			& 21\\
 & Vacuum cleaning			& 972 				& 9\\ 
 & Watching TV 			& 18648 			& 9\\  
 & Working				& 18644			& 33\\
\bottomrule
\end{tabular}
  \caption{Available segments and session in development set}
  \label{table:recscenes}
\end{table}

For the development dataset we provided the data along with reference annotations. The provided data was recorded by four out of the seven microphone arrays given the time-wise subset. The exact positions of these microphone arrays were not made available and not should not be exploited. We also provided a cross-validation setup containing four folds. Participants were encouraged to use these to make results reported on the development set uniform but it was not mandatory to use them. The daily activities (9) are shown in Table \ref{table:recscenes} along with the number of 10 second multi-channel segments and the amount of full sessions of a certain activity (e.g. a cooking session). Compared to the SINS dataset, we have combined the activity "Phone call" and "Visit" into one activity named "Social activity". The class "Absence" is referred to as not being present in the room. However, the data does contain recordings when a person is present in another room. Depending on the activity, this is noticable in the recording. The class "Other" is referred to as being present in the room but not doing any another activity in the list. All the other activities are self-explanatory. Note that the dataset is unbalanced. 

The evaluation dataset is acquired in a similar manner as the development dataset and has a similar class distribution. More statistics about this dataset will be made available after the task is finished. The dataset was provided as audio only. It contains data recorded by all seven microphone arrays on a different time-wise subset as the development set. The final evaluation will be based on the data obtained by the microphone arrays not present in the development set. The other microphone arrays are provided to give insights about the overfitting on those positions.

Participants were allowed to use external data (including pre-trained models) and data augmentation for system development. It was not allowed to use the evaluation data to train the submitted system in an (un)supervised manner.

\subsection{Baseline system}
\label{subsec:baseline}
The baseline system is intended to make it easier to participate in the task and to provide a reference performance. The system has all the functionality for dataset handling, calculating  features and models, and evaluating the results. The system is implemented in Python, primarily using the DCASE UTIL libary \cite{DCASE_UTIL} for dataset handling and feature extraction and the Keras library \cite{keras} for learning. 
\\The baseline system trains a single classifier model that takes a single channel as input. During the recording campaign, data was measured simultaneously using multiple microphone arrays each containing 4 microphones. Hence, each domestic activity is recorded as many times as there were microphones. Each parallel recording of a single activity is considered as a different example during training. The learner in the baseline system is based on a Neural Network architecture using two convolutional and one dense layer. As input, log mel-band energies are provided to the network. The features extracted from each microphone channel are treated as seperate examples. In the prediction stage a single outcome is computed for each microphone array by averaging the 4 model outcomes (posteriors) that were computed by evaluating the trained classifier model on all 4 microphones. The features are calculated in frames of 40 ms with 50 \% overlap, using 40 mel bands covering a frequency range from 50 tot 8000 Hz. An overview of the Neural Network architecture is shown in Figure \ref{fig:nn}. It uses an input size of 40x501 which are the log-mel frames in a total duration of 10 s. The first convolutional layer has 32 filters with a kernel size of 40x5 with a stride of one, so therefore convolution is only performed over the time axis. Subsampling is then performed by Max Pooling by a factor 5. The resulting feature map of 32x99 is then provided to a second convolutional layer that has 64 filters with a kernel size of 32x3 and a stride of one. Subsequentially, this is subsampled using Max Pooling by a factor 3. After each convolutional layer Batch Normalization and ReLU activation is used. The resulting output, a feature vector of 64 coefficients, is then provided to a Fully Connected (FC) layer of 64 neurons with ReLU activation. The output layer consists of 9 neurons representing the output classes with Softmax activation. For regularization we have used Dropout (20\%) between each layer. The network is trained using Adam optimizer with a learning rate of 0.0001. The used batch size is 256 segments which results in a total of 1024 examples provided to the learner given that each segment contains 4 channels of audio. On each epoch, the training dataset is randomly subsampled so that the number of examples for each class match the size of the smallest class. The performance of the model is evaluated every 10 epochs, of 500 in total, on a validation subset (30\% subsampled from the training set). The model with the highest score is used as the final model. As a metric the macro-averaged F\textsubscript{1}-score is used, which is the mean of the class-wise F\textsubscript{1}-scores.

\begin{table}[b]
\centering
\begin{tabular}{cccc} 
\toprule
Activity & F\textsubscript{1}-score dev. & F\textsubscript{1}-score eval.\\
\midrule
 Absence 				& 85.41\% 			&NA\\ 
 Cooking				& 95.14\% 			&NA\\  
 Dishwashing   			& 76.73\% 			&NA\\ 
 Eating 				& 83.64\%			&NA\\
 Other 				& 44.76\% 			&NA\\  
 Social Activity 			& 93.92\% 			&NA\\
 Vacuum cleaning			& 99.31\%			&NA\\ 
 Watching TV 			& 99.59\% 			&NA\\  
 Working				& 82.03\%			&NA\\
 \hdashline
 \textbf{Average}			& \textbf{84.50$\pm$0.8\%}	&NA\\
\bottomrule
\end{tabular}
  \caption{Performance on the development set}
  \label{table:resdev}
\end{table}

\section{Baseline system results}
\label{sec:baseres}

Table \ref{table:resdev} presents the results for the baseline system on the development set using the provided cross-validation folds.  The performance on the evaluation set will be released when the results of task are made public. Regarding the results on the development set, the system was trained and tested five times to provide an estimate on the variance related to random weight initializations and a random validation split. On average the model has a performance of 84.50\%. Class-wise performances vary from 44.76\% up to 99.59\%. Worst performing classes are Other and Dishwashing, while the best performing ones are Vacuum cleaning, Watching TV and Cooking.

\section{Conclusions}
\label{sec:concl}
In this paper we introduced the setup of the DCASE2018 Task 5 challenge, a task primarily concerned with systems exploiting both spectral and spatial information indepent of sensor location using multi-channel audio. The dataset used for the task offers recordings of domestic activities in a single home environment \cite{DevSet,EvalSet}. A baseline system was introduced based on a Neural Network architecure using two convolutional layers and a dense layer. Results were reported on this dataset using the provided publicly available baseline system \cite{DCASE2018Task5baseline} showing a macro-averaged F\textsubscript{1}-score of 84.5\%.

\section{Note}
\label{sec:note}
More information on the evaluation dataset and the performance will be made available when the task results are made public.

\section{ACKNOWLEDGMENT}
\label{sec:ack}

Thanks to Steven Lauwereins, Bart Thoen, Mulu Weldegebreal Adhana, Henk Brouckxon and Bertold Van den Bergh to their contribution of acquiring the SINS dataset \cite{Dekkers2017}.

\bibliographystyle{IEEEtran}
\bibliography{refs}

\end{sloppy}
\end{document}